\newcommand{\ket}[1] { | {#1} \rangle }
\newcommand{\bra}[1] { \langle {#1} | }

\newcommand{\PsiPlusBra} { \bra{\Psi^{+}} }
\newcommand{\PsiPlusKet} { \ket{\Psi^{+}} }

\newcommand{\rhoHat}{\hat{\rho}}

\newcommand{\Eav}{ E_{\text{av}} }
\newcommand{\EF}{ E_{F} }
\newcommand{\pureStateSet}{ \{ \ket{\psi_i} \} }
\newcommand{\decompositionSet}{ \{ p_i, \ket{\psi_i} \} }

\newcommand{\psiTildeKetI}{ \ket{\widetilde{\psi}_i} }
\newcommand{\psiTildeBraI}{ \bra{\widetilde{\psi}_i} }

\newcommand{\psiTildeKetJ}{ \ket{\widetilde{\psi}_j} }
\newcommand{\psiTildeBraJ}{ \bra{\widetilde{\psi}_j} }

\newcommand{\rhoI}{ \hat{\rho}_{i} }
\newcommand{\rhoIA}{ \hat{\rho}_{i}^{A} }
\newcommand{\rhoJA}{ \hat{\rho}_{j}^{A} }
\newcommand{\rhoIAEigenvalue}{ \lambda_{\alpha i}^{A} }
\newcommand{\rhoIAEigenket}{ \ket{\alpha_i} }

\newcommand{\rhoIANaturalLogRhoIA}{ \rhoIA \text{ln} \rhoIA }

\newcommand{\dByDEpsilon}{ \frac{d}{d\epsilon} }
\newcommand{\dByDEpsilonArg}[1]{ \frac{d#1}{d\epsilon} }

\newcommand{\logTwo}{ \text{log}_{2} }
\newcommand{\logTwoE}{ \text{log}_{2} e }
\newcommand{\traceB}{ \text{Tr}_{B} }
\newcommand{\identityMatrix}{ \textbf{I} }
\newcommand{\nullMatrix}{ \textbf{0} }

\documentclass[aps,twocolumn,showpacs,preprintnumbers,amsmath,amssymb]{revtex4}

\usepackage{graphicx}
\usepackage{dcolumn}
\usepackage{bm}

\begin{document}

\title
{
	An efficient numerical method for calculating the entanglement of formation
	of arbitrary mixed quantum states of any dimension
}

\author{J. R. Gittings}
	\email{joe.gittings@ucl.ac.uk}
	\homepage{http://www.cmmp.ucl.ac.uk/~jrg/}
\author{A. J. Fisher}
	\email{andrew.fisher@ucl.ac.uk}
	\homepage{http://www.cmmp.ucl.ac.uk/~ajf/}
\affiliation{Department of Physics and Astronomy, University College London, Gower St, London WC1E 6BT, UK}

\date{\today}

\begin{abstract}

\textbf{It has been brought to our attention that this eprint duplicates earlier research by K. Audenaert et al: see quant-ph/0006128 or PRA 64 052304.}
We present a conjugate gradient method for calculating the entanglement of formation of arbitrary mixed quantum states of any dimension and with any bipartite division of the Hilbert space. The development of the gradient used by the algorithm, its implications for the number of states required in the optimal decomposition, and the way that conjugate gradient minimization has been adapted for this particular problem are outlined. We have found that the algorithm exhibits linear convergence for general mixed states, and that it correctly reproduces the known results for pairs of qubits and for isotropic states. The results of an example application of the code are discussed: calculating the entanglement of formation of a $\ket{\Psi^{+}}$ Bell state of two qutrits when one of those qutrits is subject to various decoherence channels. The results for qutrits are contrasted with those for qubits: for the types of decoherence considered here, qutrit entanglement appears to be more robust than qubit entanglement.

\end{abstract}


\maketitle

\paragraph*{Introduction.}

The entanglement of formation $\EF$ characterizes the amount of entanglement present in a mixed state $\rhoHat$. The definition of $\EF$ is known: it is the average entanglement of the pure states $\pureStateSet$ in the decomposition of $\rhoHat$ minimized over all possible decompositions $\decompositionSet$ of $\rhoHat$. However, except for the special case of two qubits \cite{woottersPhysRevLett1998}, no formula for $\EF$ has been found, nor a prescription for obtaining the decomposition which realizes the minimum.

In this Letter, we describe the development of a gradient for the average entanglement $\Eav$, and a conjugate gradient algorithm that uses it to calculate $\EF$ numerically. The algorithm successfully minimizes the average entanglement of random two qutrit ($3 \times 3$) mixed states within a few hours on an ordinary desktop computer. The algorithm has been tested against the known result for general states of two qubits, and against exact results for $\EF$ for the special class of `isotropic' mixed states of two qutrits. Although the examples discussed in this paper involve a division of the Hilbert space into subspaces of equal dimension, the algorithm works equally well for any division of the Hilbert space. For example, it could be used to study the entanglement between a qutrit and quqit (d=4 system) whose joint state is mixed.

\paragraph*{The entanglement of formation.}

A particular pure state decomposition of a mixed state $\hat{\rho}$ shared between subsystems $A$ and $B$ can be written
\begin{eqnarray}
	\hat{\rho} &=& \sum_i p_{i} \rhoHat_{i} = \sum_i p_{i} \ket{\psi_i} \bra{\psi_i} = \sum_i \psiTildeKetI \psiTildeBraI
\end{eqnarray}
where the
$\{ \psiTildeKetI \}$ are subnormalized states defined by $\psiTildeKetI := \sqrt{p_{i}} \ket{\psi_i}$
\cite{nielsenAndChuangPage103}.
The average entanglement of the decomposition, in ebits, is
\begin{eqnarray}
	E_{av} ( \{ \psiTildeKetI \} ) &=& \sum_{i} p_{i} E_{i}
	\nonumber \\
	&=& -\text{log}_{2} e \sum_{i} p_{i} \text{Tr}_{A} [ \rhoIA \ln \rhoIA ]
\end{eqnarray}
where $E_{i}$ is the entanglement of $\ket{\psi_i}$. For convenience, the derivation works with natural logs. The entanglement of formation is defined as the minimum of $E_{av}$ over all possible decompositions of $\rhoHat$:
\begin{eqnarray}
	\EF(\rhoHat) := \min_{ \{ \psiTildeKetI \} } E_{av}
\end{eqnarray}
The decomposition which gives the minimum is described as `optimal'. It is possible to make unitary transformations between different decompositions of a given mixed state \cite{nielsenAndChuangPage103}:
\begin{eqnarray}
	\label{eqn:unitaryFreedom}
	\psiTildeKetI \rightarrow \sum_j U_{ij} \psiTildeKetJ
\end{eqnarray}
Although these decompositions all generate the same mixed state, they have different values of the average entanglement.

\paragraph*{A gradient for the average entanglement.}

Now write
\begin{eqnarray}
	U = \text{exp} (i \epsilon \theta)
\end{eqnarray}
where $\epsilon$ is a real parameter, and $\theta$ is a Hermitian matrix. All derivatives are evaluated at $\epsilon=0$. The gradient of $\rhoIA$ is easily obtained:
\begin{eqnarray}
	\frac{ d\rhoIA }{ d\epsilon } &=&
		\frac{1}{p_i} \dByDEpsilon \biggl( \text{Tr}_{B} (\psiTildeKetI \psiTildeBraI) \biggr)
		- \frac{1}{p_i^{2}} \frac{dp_i}{d\epsilon} \text{Tr}_{B} (\psiTildeKetI \psiTildeBraI)
\end{eqnarray}

The gradient of $\text{Tr}_{A} [ \rhoIA \text{ ln } \rhoIA ]$ requires a little more work. If $\alpha$ labels an eigenvalue $\rhoIAEigenvalue$ of $\rhoIA$, and $\rhoIAEigenket$ is the corresponding eigenket,
\begin{eqnarray}
	\label{eqnDByDEpsilonSRhoIA}
	\dByDEpsilon \text{Tr}_{A} [ \rhoIA \text{ ln } \rhoIA ]
		&=& \dByDEpsilon \sum_{\alpha} \rhoIAEigenvalue \text{ ln } \rhoIAEigenvalue
		\nonumber \\
	&=& \text{Tr}_{A} \biggl( \text{ ln } \rhoIA \dByDEpsilonArg{\rhoIA} \biggr)
\end{eqnarray}
where we have made use of the invariance of $\text{Tr}(\rhoIA)$ and the Hellman-Feynman theorem
\cite{ehrenfest1927,hellmann1937,feynman1939}.
Now we can calculate the promised expression for the gradient:
\begin{eqnarray}
	\dByDEpsilonArg{E_{av}} &=&
		-\logTwoE \sum_{i} \biggl( \dByDEpsilonArg{p_i} \text{Tr} [\rhoIANaturalLogRhoIA] +
		p_{i} \text{Tr} [\text{ln } \rhoIA \dByDEpsilonArg{\rhoIA}] \biggr) 
		\nonumber \\
	&=& -\logTwoE \sum_{i} \text{Tr}_{A} \biggl[ \text{ln} \rhoIA \dByDEpsilon \text{Tr}_{B} \psiTildeKetI \psiTildeBraI \biggr]
\end{eqnarray}
The result can be further simplified by calculating the gradient of
$p_{i} \rhoIA = \traceB (\psiTildeKetI \psiTildeBraI)$ and making use of equation (\ref{eqn:unitaryFreedom}):
\begin{eqnarray}
	U &=& e^{i \epsilon \theta}
	\therefore \dByDEpsilon U_{ij} \biggl|_{\epsilon=0} = i \theta_{ij}
	\nonumber \\
	\label{rgd_intermediate}
	\therefore
	\dByDEpsilon{(p_i \rhoI)} \biggl|_{\epsilon=0} &=&
		\dByDEpsilon \biggl( \sum_{j} U_{ij} \psiTildeKetJ \biggr) \psiTildeBraI +
		\nonumber \\
		&& \psiTildeKetI \dByDEpsilon \biggl( \sum_{j} \psiTildeBraJ (U^{\dag})_{ji} \biggr)
		\nonumber \\
		&=& i \sum_{j} ( \theta_{ij} \psiTildeKetJ \psiTildeBraI
			- \theta_{ji} \psiTildeKetI \psiTildeBraJ )
\end{eqnarray}
\begin{eqnarray}
	\therefore
	\dByDEpsilonArg{E_{av}} &=&
		- \sum_{i} \text{Tr}_{A} \biggl[ \logTwo \rhoIA \biggl\{ 
		i \sum_{j} ( \theta_{ij} \traceB \psiTildeKetJ \psiTildeBraI
		\nonumber \\
		&&
		- \theta_{ji} \traceB \psiTildeKetI \psiTildeBraJ )
		\biggr\} \biggr]
	\nonumber \\
	\label{eqn:final_gradient}
	&=& -i \sum_{i,j} \theta_{ij} \text{Tr}_{A} \biggl[
		\biggl( \logTwo \rhoIA - \logTwo \rhoJA \biggr) 
		\traceB \psiTildeKetJ \psiTildeBraI
		\biggr]
	\nonumber \\
	&=& \sum_{i,j} \theta_{ij} g_{ji}
\end{eqnarray}
where $\textbf{g}$ is the matrix of gradient elements with respect to the space of generators $\{ \theta \}$.

\paragraph*{Number of states required in the optimal decomposition.}

Suppose we add to our set of states a new one, sub-normalized to zero, and allow unitary transformations among this new, enlarged, set. Such transformations automatically leave the density matrix invariant, but it seems possible that they might reduce the average entanglement.  However, equation~(\ref{eqn:final_gradient}) shows that they do not, at least to first order, since the trace over B on the RHS will be zero if either $\psiTildeKetI$ or $\psiTildeKetJ$ is sub-normalized to zero.  Therefore no `padding' with zero-norm states is required to produce the correct result. This strongly suggests that the number of states needed is equal to the rank of the density matrix $\rhoHat$, instead of the previously accepted bound of the square of the dimension of $\rhoHat$.

\paragraph*{Existence of local and global minima}

In general, a conjugate gradient procedure will converge to a local, rather than a global, minimum. However, Prager \cite{prager2001}, has shown that any local minimum of $\EF$ is also a global minimum. We have encountered points (presumably local maxima or inflexion points) other than the optimal decomposition at which the gradient is zero: when this occurs we find we can restart the conjugate gradient algorithm after a small number of random moves.

\paragraph*{Using the gradient to minimize the average entanglement.}

The code described here uses the conjugate gradient algorithm to minimize $\Eav$ with respect to the space of all possible unitary transformations between some initial decomposition of $\rhoHat$ (the choice of which is entirely arbitrary) and the current decomposition. The initial decomposition is chosen without prejudice to be the `eigenstate' decomposition of $\rhoHat$, ie. consisting of pure states which are the eigenvectors of $\rhoHat$ and probabilities which are the eigenvalues. We write
$\textbf{U}=\text{exp}(i\textbf{H})$
and move through the space of Hermitian matrices $\{ \textbf{H} \}$ by constructing conjugate gradient moves from the gradient information provided by $\textbf{g}$. The initial unitary transformation is the identity $\identityMatrix$, whose corresponding $\textbf{H}$ is the null matrix $\nullMatrix$.

The standard formulation of the conjugate gradient algorithm \cite{recipesConjugateGradientDescription} operates upon vectors. Therefore the point in \textbf{H}-space used is the `flattened' vector consisting of the unique elements of $\textbf{H}$. The gradient matrix is flattened in the same way. Conjugate directions are chosen according to the Fletcher-Reeves-Polak-Ribiere algorithm \cite{polakBookFRPRMN}. The gradient information is also made use of by the modified version of Brent's method \cite{brentBookDLINMIN} that performs the line minimizations. The end result of the algorithm is the unitary transformation which takes us from the `eigenstate' decomposition to the optimal decomposition. An heuristic test for convergence is performed after the algorithm terminates by running Monte Carlo against the final decomposition, using an exponentially wide range of step sizes in random directions through unitary transformation space: we fail to find further steps down to within the target precision. This does not prove that the optimal decomposition has been obtained, but is strongly suggestive of that conclusion. Note that the gradient for $\Eav$ is constructed with respect to small unitary transformations of the current decomposition, and therefore the unitarity of $U$ is preserved to machine precision throughout the execution of the conjugate gradient algorithm.

\paragraph*{Performance against two-qubit mixed states}

The algorithm's estimate of $\EF$ converges with the Wootters $\EF$ to 14 decimal places typically in less than 100 iterations.

\paragraph*{Performance against two-qudit isotropic mixed states}

An isotropic mixed state is a convex mixture of the maximally entangled Bell state $\PsiPlusKet$ and the maximally mixed state $\rho_{I} = \identityMatrix$:
\begin{eqnarray}
	\rhoHat := \frac{1-F}{d^2 - 1} (\identityMatrix - \PsiPlusKet \PsiPlusBra) + F \PsiPlusKet \PsiPlusBra
\end{eqnarray}
Unlike for more general mixed states, a formula for $\EF$ exists for isotropic states of two qudits \cite{terhal2000}. These states are thus well-suited as a check for the results of the algorithm. In the case of two-qubit isotropic states, a particular difficulty arises which is that $\Eav$ generally has a stationary point at the eigenvector decomposition. It is thus necessary to move the initial decomposition away from this stationary point with a few random Monte Carlo moves, after which the conjugate gradient algorithm successfully obtains the correct minimum. For two-qutrit isotropic states we observe perfect correspondence between the results of the algorithm and the value given by the formula.

\paragraph*{Performance against two-qutrit random mixed states.}

\begin{figure*}
	\caption
	{
		\label{fig:two_qutrit_convergence}
		Convergence of min($\Eav$) for a random two-qutrit mixed state using conjugate gradient algorithm.
	}

	\includegraphics[width=3in]{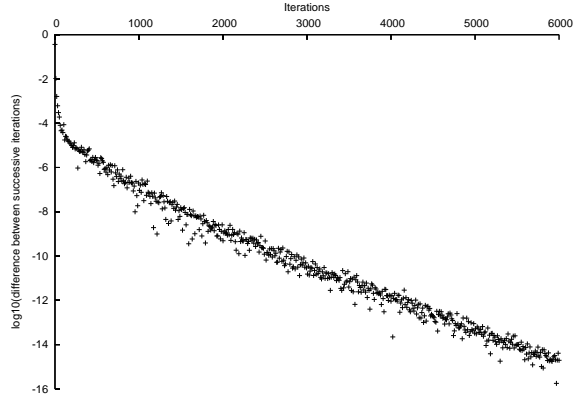}
\end{figure*}

The convergence behaviour when minimizing a sample random general mixed state of two qutrits (ie. $\text{dim}(\mathcal{H})=9$) is shown in Figure \ref{fig:two_qutrit_convergence}. It can be seen that convergence is linear, with approximately 600 iterations required per significant figure.  We have found a few examples of states with highly degenerate eigenvalue spectra (e.g.\ isotropic states) for which convergence may be slower than linear.

\paragraph*{Modelling locally depolarized Bell states of two qutrits.}

An example of a previously intractable yet conceptually simple problem is local depolarization of one qudit involved in an entangled state. Consider a $\PsiPlusKet$ state of two systems of dimension $d$ (qudits), defined by
\begin{eqnarray}
	\PsiPlusKet := \frac{1}{\sqrt{d}} \sum_{i} \ket{ii}.
\end{eqnarray}
If one qudit - say qudit $A$ - decoheres with a certain probability $p$, the entanglement of the overall state is reduced in a non-trivial manner. This corresponds to a physical situation in which one party (Bob) prepares two qudits in an entangled state, then passes one qudit to Alice via a noisy channel.  How much entanglement do Alice and Bob now share?

\begin{figure*}
	\caption
	{
		\label{fig:qubit_local_decoherence}
		Entanglement of formation (ebits) of $\PsiPlusKet$ of two qubits vs probability
		of qubit $A$ depolarizing through various channels
		(calculated using Wootters formula).
	}

	\includegraphics[width=6.2in]{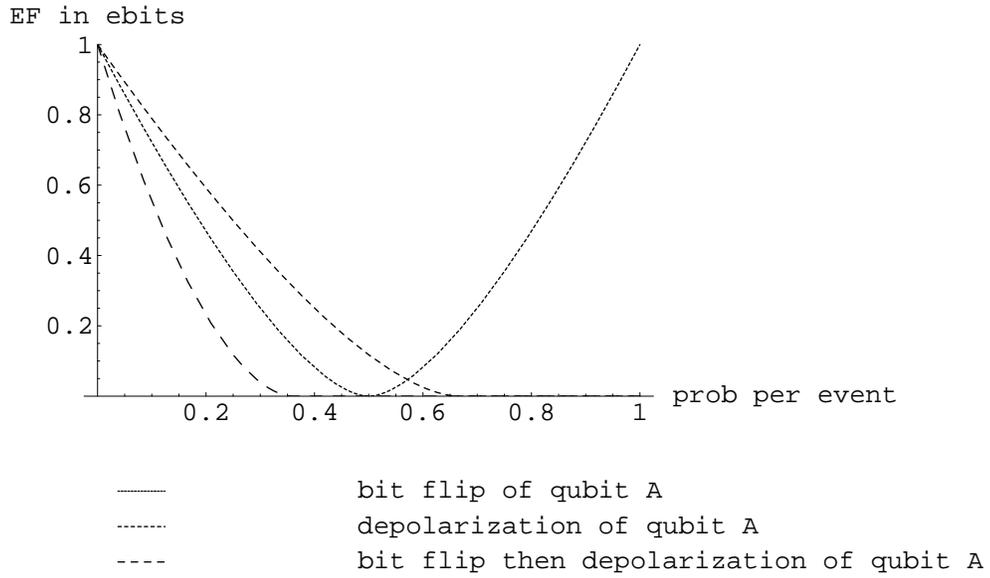}
\end{figure*}

In this exercise, three possible types of decoherence are considered: the bitflip channel, the depolarizing channel, and both channels in succession with the same probability. (As the operations commute in this case, the order of bitflip and depolarization operations in the last case is irrelevant.) The results for a $\PsiPlusKet$ state of two qubits can be readily calculated using the Wootters formula. They are shown in Figure \ref{fig:qubit_local_decoherence}. For the depolarizing channel, the results are consistent with the well-known separability of mixed states in the neighbourhood of the maximally mixed state \cite{braunsteinPhysRevLett1999}.

\begin{figure*}
	\caption
	{
		\label{fig:qutrit_local_decoherence}
		Entanglement of formation (etrits) of $\PsiPlusKet$ of two qutrits vs probability of
		qutrit $A$ depolarizing through various channels
		(calculated using conjugate gradient algorithm).
	}

	\includegraphics[width=3in]{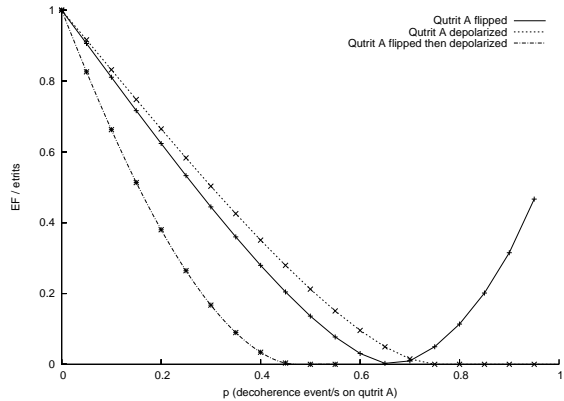}
\end{figure*}

To perform the same exercise for a $\PsiPlusKet$ state of two qutrits ($d=3$), the bitflip channel is extended to qutrits by using an operator sum representation consisting of 
$\sqrt{1-p} \textbf{I}, \sqrt{p/2} \textbf{X}_{up}, \sqrt{p/2} \textbf{X}_{down}$,
where $p$ represents the probability of a flip and the latter two operators respectively raise and lower a basis state. The depolarizing channel is extended to qutrits using the operator-sum representation in \cite{barg2002} with $d=3$.  The results of applying these operations (tensored with the identity on qutrit $B$, which does not decohere) are shown in Figure \ref{fig:qutrit_local_decoherence}.  The results are plotted in etrits
(1 etrit $= \text{log}_2(3)$ ebits)
to facilitate comparison with the results for qubits. (Note that for the depolarization-only case, the results can also be obtained using the formula for $\EF$ for isotropic states - we have found they are the same). They are similar to those for qubits, however it can be seen that for all types of channel, the proportionate reduction in entanglement at a given probability is lower for qutrits. It thus appears that qutrit entanglement is more robust with respect to these types of decoherence than qubit entanglement. This conclusion has potential significance for the design of quantum computers and communication channels.

\paragraph*{Acknowledgements.}

We would like to thank V. Vedral and S. Bose for many helpful discussions on this and related topics, and EPSRC for financial support.

\bibliographystyle{unsrt}
\bibliography{mineav}

\begin{thebibliography}{10}

\bibitem{woottersPhysRevLett1998}
W.~K. Wootters.
\newblock {\em Phys. Rev. Lett.}, 80:2245--2248, 1998.

\bibitem{nielsenAndChuangPage103}
M.~A. Nielsen and I.~L. Chuang.
\newblock {\em Quantum computation and quantum information}, page 103.
\newblock Cambridge University Press, 1st edition, 2000.

\bibitem{ehrenfest1927}
P.~Ehrenfest.
\newblock {\em Z. Phys.}, 45:455, 1927.

\bibitem{hellmann1937}
H.~Hellmann.
\newblock {\em Einfuhrung in die Quantenchemie}.
\newblock Deuticke, Leipzig, 1937.

\bibitem{feynman1939}
R.~P. Feynman.
\newblock {\em Phys. Rev.}, 56:340, 1939.

\bibitem{prager2001}
T.~Prager, 2001.
\newblock quant-ph/0106030.

\bibitem{recipesConjugateGradientDescription}
W.~H.~Press et~al.
\newblock {\em Numerical recipes in C++, the art of scientific computing},
  pages 424--428.
\newblock Cambridge University Press, 1st edition, 2002.

\bibitem{polakBookFRPRMN}
E.~Polak.
\newblock {\em Computational methods in Optimization}.
\newblock New York: Academic Press, 1971.

\bibitem{brentBookDLINMIN}
R.~P. Brent.
\newblock {\em Algorithms for minimization without derivatives}.
\newblock Prentice-Hall, 1973.

\bibitem{terhal2000}
B.~M. Terhal and K.~G.~H. Vollbrecht, 2000.
\newblock quant-ph/0005062.

\bibitem{braunsteinPhysRevLett1999}
S.~L.~Braunstein et~al.
\newblock {\em Phys. Rev. Lett.}, 83:1054--1057, 1999.

\bibitem{barg2002}
A.~Barg.
\newblock {\em IEEE Trans. Inform. Theory}, 48, 2002.

\end{thebibliography}

\end{document}